\documentclass[aps,pre,preprint,showpacs,amsmath,amssymb]{revtex4}

\usepackage{graphicx}
\usepackage{dcolumn}
\usepackage{bm}
\usepackage{verbatim}
\usepackage{setspace}
\usepackage{ulem}

\renewcommand{\a}{\displaystyle}

\begin{document}

\title{Mean survival times of absorbing triply periodic minimal surfaces}

\author{Jana Gevertz$^1$ and S. Torquato$^{1,2,3,4,5}$}
\email{torquato@princeton.edu}

\affiliation{$^1$Program in Applied and Computational Mathematics,
Princeton University, Princeton New Jersey 08544, USA}

\affiliation{$^2$Department of Chemistry, Princeton University,
Princeton, New Jersey 08544, USA}

\affiliation{$^3$Princeton Center for Theoretical Science,
Princeton University, Princeton, New Jersey 08544, USA}

\affiliation{$^4$Princeton Institute for the Science and Technology of Materials,
Princeton University, Princeton, New Jersey 08544, USA}

\affiliation{$^5$School of Natural Sciences, Institute for
Advanced Study, Princeton, New Jersey 08540, USA}

\date{\today}

\begin{abstract} 
 Understanding the transport properties of a
porous medium from a knowledge of its microstructure is a problem
of great interest in the physical, chemical and biological sciences.
Using a first-passage time method, we compute the mean survival time $\tau$ 
of a Brownian particle among perfectly absorbing traps 
for a wide class of  triply-periodic porous media, including minimal 
surfaces. We find that the porous medium with an interface that is the Schwartz 
P minimal surface maximizes the mean survival time among this class. This adds 
to the growing evidence of the multifunctional optimality of this bicontinuous 
porous medium.  We conjecture that the mean survival time (like the fluid 
permeability) is maximized for triply periodic porous media with a simply 
connected pore space at porosity $\phi=1/2$ by the structure that globally 
optimizes the specific surface. We also compute pore-size statistics
of the model microstructures in order to ascertain the validity
of a ``universal curve" for the mean survival time for these porous media.
This represents the first nontrivial statistical characterization of triply
periodic minimal surfaces.
\end{abstract}


\maketitle

\section{INTRODUCTION}

Fluid-saturated porous media are ubiquitous in nature (e.g., geological and
biological media) and in synthetic situations (e.g., filters, cements and
foams)~\cite{torquato02,Sa03}.  Understanding the transport properties of a
porous medium from a knowledge of its microstructure is a subject that spans
many fields~\cite{torquato02,Sa03,Co96,Ka02,Ol04,Ha07}. In particular, physical 
phenomena involving simultaneous diffusion and reaction in porous media abound 
in the physical and biological sciences~\cite{torquato02,Sa03,Be93}. 
Considerable attention has been devoted to instances in which diffusion occurs 
in the pore region of the porous medium with a ``trap'' region 
whose interface can absorb the diffusing species via a surface reaction.  
Examples are found in widely different processes, such as heterogeneous catalysis, 
fluorescence quenching, cell metabolism, ligand binding in proteins, migration of 
atoms and defects in solids, and crystal growth \cite{torquato02}. A key parameter 
in such processes is the {\it mean survival time} $\tau$, which gives the average 
lifetime of the diffusing species before it gets trapped. 

\begin{figure}[ht!]
 \includegraphics[width=4.5in]{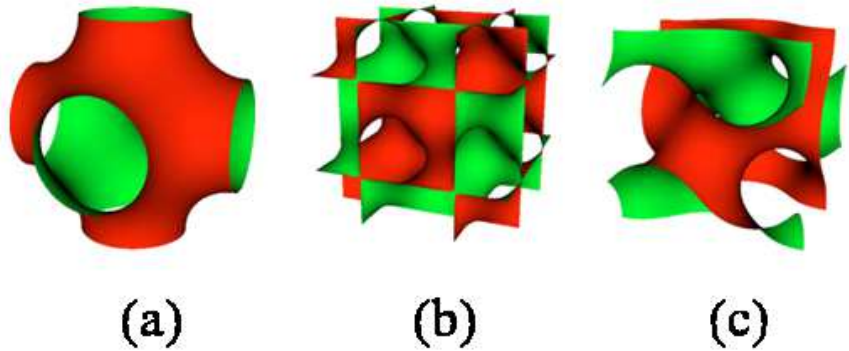} 
 \caption{\label{fig:1} (Color online) Unit cells of three different minimal 
 surfaces. (a) Schwartz  P surface, (b) Schwartz D surface, (c) 
 Schoen G surface.  Image is adapted from Ref. \cite{jung05}.}
\end{figure}

It is noteworthy that while there has been a significant amount of progress
made on the determination of the structures that optimize a 
variety of transport and mechanical properties of porous media \cite{torquato02,hashin,kohn}, 
there have been no studies that have attempted to find
the optimal isotropic structures for the mean survival time $\tau$ \cite{To04}. 
It has been recently demonstrated that triply periodic two-phase bicontinuous 
composites with interfaces that are the Schwartz P and D minimal surfaces are 
not only geometrically extremal, but are also extremal for simultaneous 
transport of heat and electricity~\cite{torquato02b,compet,Si07}. A minimal 
surface is one that is locally area minimizing i.e., every point has zero mean 
curvature. Triply periodic minimal surfaces are minimal surfaces that are 
periodic in all three coordinate directions. An important subclass of 
triply-periodic minimal surfaces are those that partition space into two 
disjoint but intertwining regions that are simultaneously continuous (i.e., 
bicontinuous)~\cite{anderson90,jung07}.  Examples of such surfaces include the 
Schwartz primitive (P), the Schwartz diamond (D), and the Schoen gyroid (G) 
surfaces [see Fig. \ref{fig:1}]; each disjoint region has volume 
fraction equal to 1/2. Such triply periodic minimal surfaces arise in a variety 
of systems, including  nanocomposites \cite{nano},  micellar materials 
\cite{micelle}, block copolymers \cite{block} as well as lipid-water systems 
and certain cell membranes \cite{mem1,mem2,mem3,mem4}.

The multifunctionality of such two-phase systems has been further established 
by showing that they are also extremal when a competition is set up between 
the effective bulk modulus and electrical (or thermal) conductivity of the 
bicontinuous composite~\cite{torquato04}. Jung and Torquato~\cite{jung05} 
also explored the multifunctionality of the three minimal surfaces shown in Fig. 
\ref{fig:1} with respect to Stokes (slow viscous) flow.  The simulations 
conducted by Jung and Torquato revealed that the Schwartz P porous medium has 
the largest fluid permeability $k$ among a class of structures studied with 
porosity $\phi=1/2$. Further, the fluid permeability was found to be inversely 
proportional to the specific surface $s$ (interface area per unit volume). This 
led the authors to conjecture that the maximal fluid permeability for a triply 
periodic porous medium with a simply connected pore space \cite{simply} at a porosity  
$\phi$=1/2 is achieved by the structure that globally minimizes the specific 
surface~\cite{jung05}.

In this manuscript, we explore whether the mean survival time $\tau$ of a 
Brownian particle among perfectly absorbing traps is maximized by the Schwartz 
P structure among a wide class of  triply-periodic porous media based 
on the class of six models studied in Ref. \cite{jung05}.  Further, we go 
beyond considering just a representative medium from each class of models and 
explore several of the models over a {\it range} of parameter values, as will 
be further explained below. This inquiry is also motivated by a 
cross-property bound that rigorously links the fluid permeability $k$ to the 
mean survival time $\tau$~\cite{torquato90}:
\begin{equation}
 k \le \tau.
\label{k-tau}
\end{equation}
It should be noted that the inequality of (\ref{k-tau}) becomes an equality for 
transport in parallel cylindrical tube bundles of arbitrary cross-sectional 
geometry \cite{torquato90,tube}.  It is also a relatively tight bound (when 
appropriately scaled) for transport around distributions of inclusions 
\cite{torquato90}.  It is clear that the permeability $k$ will be maximized
for a simply connected pore space (e.g., the presence of dead ends 
are undesirable because they would not contribute to the flow).
The inequality (\ref{k-tau})  and the results  of Ref. 
\cite{jung05} suggest that $\tau$ is maximized by the same simply connected microstructure that 
maximizes $k$ for $\phi=1/2$. 

We will also test whether our calculations for $\tau$ collapse on to a 
``universal" scaling relation for the mean survival time~\cite{torquato97}.
This requires us to compute the pore-size density functions for the triply 
periodic surfaces considered in this paper.

In Sec. II, we define terminologies and give a precise statement of the 
problem. In Sec. III, we describe the first-passage time technique that we 
utilize to compute the mean survival time for a wide class of triply periodic 
porous \textbf{media} that are generally bicontinuous.
Section IV reports our finding for $\tau$.
In Sec. V, we discuss a universal scaling relation for $\tau$,
report pore-size statistics for the triply periodic porous media
studied here, and ascertain the applicability of the universal
curve for these structures. Finally, in Sec. VI, 
we make concluding remarks and discuss the ramifications
of our results.

\section{Definitions and Problem Statement}

The mean survival time $\tau$ arises in steady-state diffusion of reactants in 
a trap-free pore region $\mathcal{V}_1$ with diffusion coefficient $D$ among 
static traps with a unit rate of production of the reactants per unit pore 
volume~\cite{torquato02}.  When the reactants come in contact with the 
pore-trap interface  $\partial\mathcal{V}$, they get absorbed. Using 
homogenization theory~\cite{torquato02}, it has been shown that $\tau$,
the average time traveled before a diffusing particle gets trapped, is given 
by:
\begin{equation}
 \tau = \frac{\a \langle u \rangle}{\a \phi D},
\end{equation}
where $u(\mathbf{r})$ is the scaled concentration field of reactants, which 
satisfies the diffusion equation
\begin{equation}
 \Delta u = -1 \quad \text{ in }\mathcal{V}_1
\end{equation}
\begin{equation}
 u = 0 \quad    \text{ on } \partial\mathcal{V},
\end{equation}
and $\Delta$ is the Laplacian operator.

\begin{figure}[ht!]
 \includegraphics[width=4.in]{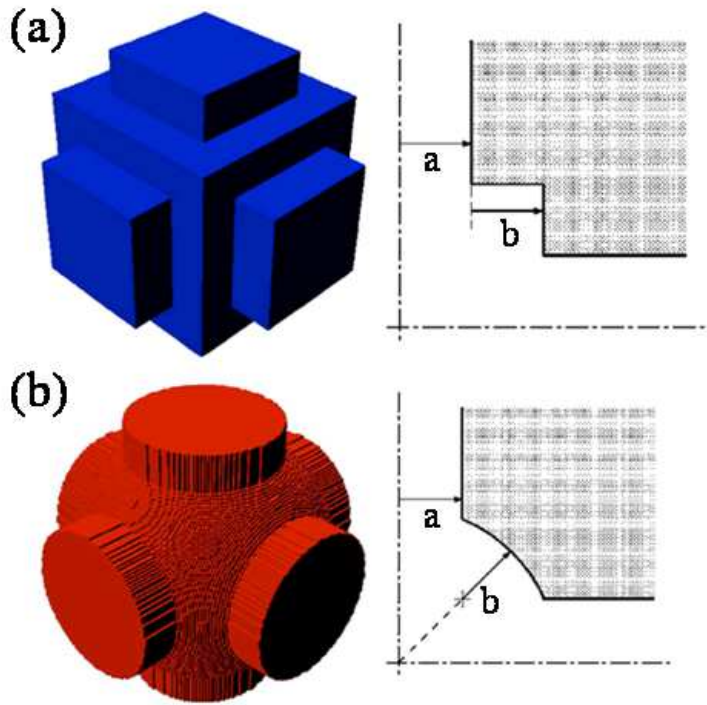}
 \caption{\label{fig:2} (Color online) Unit cells of two different pore-channel 
 models. (a) Cubic pore-square channel and its cross-section.  
 (b) Spherical pore-circular channel and its cross section.  
 Image is adapted from Ref. \cite{jung05}.}
\end{figure}

Following Ref. \cite{jung05}, we compute $\tau$ for a wide class triply 
periodic structures: the Schwartz P, Schwartz D and Schoen G minimal surfaces 
[Fig.~\ref{fig:1}],  a cubic pore-square channel model [Fig.~\ref{fig:2}(a)], 
and a spherical pore-circular channel model [Fig.~\ref{fig:2}(b)], and an 
array of spherical traps arranged on a simple cubic lattice~\cite{To92} 
(which of course is not bicontinuous). The pore-channel models each contain 
two parameters, $a$ and $b$, that enable one to control the relative size of 
the pores (either spherical or cubic in shape) to the size of the channels 
(either circular or square cross-sections)~\cite{jung05}.  The parameter $a$ 
determines channel width/diameter, whereas $b$ determines the width/diameter 
of the pore.  This class of models is bicontinuous provided that $a>0$.
In the current investigation, we compute $\tau$ for the pore-channel 
models using a wider range of values for $a$ and $b$ than used in Ref. \cite{jung05}, while keeping $\phi$ fixed 
at 1/2.

\section{FIRST-PASSAGE TIME METHOD FOR COMPUTING MEAN SURVIVAL TIME}

The mean survival time $\tau$ for the aforementioned triply periodic porous 
media could be calculated using standard random-walk techniques 
that simulate the detailed zig-zag trajectory of a Brownian particle \cite{Le89}.
We instead compute $\tau$ for these microstructures using an efficient 
first-passage time algorithm~\cite{torquato89}. The fundamental periodic 
cell is taken to be a cube of side length unity.

\begin{figure}[ht!]
 \includegraphics[width=4.0in]{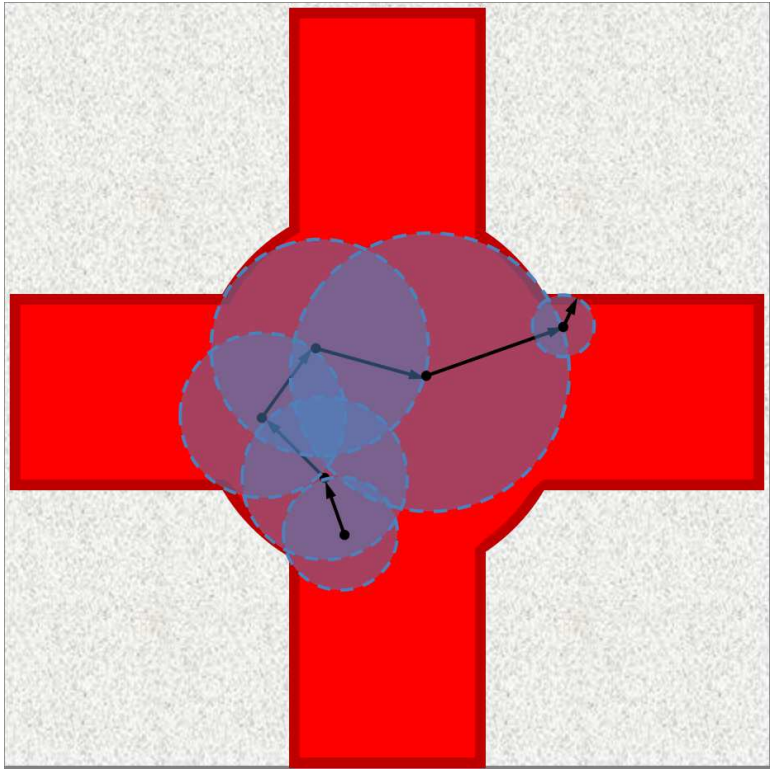}
 \caption{\label{fig:FPT} (Color online) Two-dimensional depiction of 
 continuous first-passage time method applied to the spherical 
 pore-circular channel model.  The Brownian particle walks, by jumping
to the surface of the largest possible first-passage  sphere at each
step, until 
 it gets trapped at the two-phase interface (maroon boundary layer, which 
 appears as dark grey if the image is being viewed in black and white).}
\end{figure}

The first-passage time algorithm is implemented by applying the following 
set of rules:

\begin{enumerate}
 \item Introduce a Brownian particle into a random position in the 
 trap-free phase.
 \item While the walker is sufficiently far from the 
 two-phase interface, construct the largest sphere of radius $R_i$ centered 
 at the Brownian particle that just touches the two-phase interface.
 \item In one step, the walker jumps to a random point on the 
 surface of the sphere [Fig.~\ref{fig:FPT}], with an average hitting 
 time of:
  \begin{equation}
   \bar{t} = R_i^2/(6D).
  \end{equation}
 \item Repeat steps (2)-(3) until the walker is within some prescribed 
 small distance $\delta$ (taken to be $10^{-8}$ in our simulations) from 
 the two-phase interface.
 \item Repeat steps (1)-(4) for many random walkers and calculate the 
 mean survival time from the following equation: 
  \begin{equation}
   \tau = \frac{\a \left\langle \sum_i R_i^2 \right\rangle}{\a 6D},
  \end{equation}
 where angular brackets denote an ensemble average.
\end{enumerate}

The mean survival time $\tau$ of the pore-channel models are determined 
using this algorithm.  However, since we only have  discrete (not continuous) 
representations of the minimal surfaces~\cite{jung05}, the discrete analog 
of the algorithm is utilized in these cases~\cite{torquato99}.  In the 
discrete case, first-passage cubes of length $2R_i$ are utilized instead 
of first-passage spheres, and the average time it takes to move to the 
surface of a first-passage cube is $\bar{t} \approx 0.225R_i^2/D$.

\section{RESULTS FOR THE MEAN SURVIVAL TIMES}

The value of $\tau$ for each of the 
aforementioned models is given in 
Table~\ref{tab:table1}, along with the fluid permeability and specific 
surface. All fluid permeability measurements (if available), as well as 
the specific surface measurements for the minimal surfaces, and the 
pore-channel models with $b=0$ are taken from Ref. \cite{jung05}.  We see 
that $\tau$ always obeys the rigorous bound specified by (1). We also see 
that a simply connected pore phase is a 
crucial topological feature required to achieve large mean 
survival times at a porosity $\phi=1/2$.  However, a simply connected pore 
phase is not a sufficient condition, as evidenced by the
relatively small mean survival times associated with the Schwartz D and 
Schoen G minimal surfaces. These two minimal surfaces have 
rather large specific surfaces and hence serve as efficient traps for the 
diffusing Brownian particles. Indeed, we see that the survival times are 
inversely proportional to the corresponding specific surfaces for all of the 
bicontinuous structures. The pore spaces of structures with large $\tau$ 
are expected to be simply connected and therefore it would not be 
unreasonable for the survival time to be inversely proportional to the 
specific surface in these instances. As hypothesized, we find that among 
the structures examined, the Schwartz P porous medium maximizes the mean 
survival time. To get some idea of the fluctuations about the average
$\tau$ values, we computed the associated variance $\sigma_{\tau}^2$ for trapping
for the three minimal surfaces and found that $\sigma_{\tau}^2=0.00051, 0.00013$
and $0.00005$ for the Schwartz P, Schoen G and Schwartz D surfaces,
respectively.

\begin{singlespace}
\begin{table}[ht!]
 \caption{\label{tab:table1} Mean survival time $\tau$, fluid 
 permeability $k$, and specific surface  $s$ of triply periodic 
 structures. All quantities are made dimensionless using the 
side length of the unit cell.}
  \begin{ruledtabular}
   \begin{tabular}{l|c|c|c}
    \textbf{Structure} & $\boldsymbol{\tau}$ & $\mathbf{k}$ & 
    $\mathbf{s}$ \\
     \hline\hline
      Schwartz P & 0.0173950 & 0.0034765 & 2.3705  \\
      Schoen G & 0.0093266 & 0.0022889 & 3.1284  \\
      Schwartz D & 0.0060414 & 0.0014397 &  3.9011  \\ \hline
      Cubic-pore channel  &  &  & \\
       \hspace{0.05in} ($a=0.25$; $b=0$) & 0.0139289 & 
        0.0030744 & 3.0000 \\
       \hspace{0.05in} ($a=0.1324$; $b=0.25$)  & 0.0123813 & 
        0.0005310& 3.8360  \\
       \hspace{0.05in} ($a=0.0781$; $b=0.3125$) & 0.0122259 & 
        0.0000948 & 3.9254 \\
       \hspace{0.05in} ($a=0$; $b=0.3969$) & 0.0127852 & 
        - & 3.7804 \\ \hline
      Sphere-pore channel & & & \\
       \hspace{0.05in} ($a=0.2836$; $b=0$)  & 0.0167934 & 
        0.0034596 & 2.63990 \\
       \hspace{0.05in} ($a=0.1480$; $b=0.2794$) & 0.0164995 & 
        - & 2.73649 \\
       \hspace{0.05in} ($a=0.0908$; $b=0.3633$) & 0.0161733 & 
        - & 2.93057 \\
       \hspace{0.05in} ($a=0$; $b=0.4924$) & 0.0161093 & - & 
        3.04681 \\ \hline
      Spherical trap ($\phi = 0.5$) & 0.0139640 & 0.0030591 &
       3.0780 \\
   \end{tabular}
  \end{ruledtabular}
\end{table}
\end{singlespace}

Given that both the mean survival time $\tau$ and fluid permeability
$k$ are made dimensionless with the 
side length of the unit cell, one might ask why these dimensionless
bulk properties are two to three orders of magnitude smaller than unity?
This is a well-known behavior for such transport properties of 
porous media, optimal or not \cite{torquato02,Sa03}. For example, the fluid permeability
(average of the velocity field) can be regarded to be the ``effective pore channel area of
the dynamically connected" part of the pore space. For non-simply
connected pore spaces, there generally will be regions
that contain fluid but do not actively contribute to the flow
(not dynamically connected). Moreover, because of the no-slip
condition, the velocity only becomes significantly large sufficiently
away from the pore-solid interface for general porous media.  Thus, these two effects
conspire to make  the effective area of a pore channel, i.e., the
fluid permeability $k$,  considerably less than 
the geometric pore channel sizes. (This effect is even true
for the simple case of flow in a tube; see Ref. \cite{tube}). Similar arguments apply
to the mean survival time. The perfectly absorbing boundary
condition at the pore-solid interface (i.e., zero concentration
field) means that the concentration field becomes appreciably  large sufficiently
away from the pore-solid interface, which results in a mean survival
time (average of the concentration field) that is generally several
orders of magnitude smaller than dictated by the largest pore dimensions.

\section{ Pore-Size Functions and Universal Scaling Relation for Survival Time}

\subsection{Pore-Size Function}

Porous media whose interfaces are triply periodic minimal surfaces 
apparently have remarkable macroscopic properties. Nonetheless,
these structures have yet to be statistically characterized using
nontrivial descriptors. Here we present pore-size functions
for these structures as well as the triply periodic circular channels.
This is motivated by the fact that the pore-size density function $P(\delta)$
arises in rigorous lower bounds on $\tau$ \cite{torquato02} as well as a
universal curve for $\tau$ \cite{torquato97}, described below.
The quantity  $P(\delta)d\delta$ gives the probability that a randomly 
chosen point in the pore region lies at a distance between $\delta$ and 
$\delta + d\delta$ from the nearest point on the 
interface~\cite{torquato02}.

\begin{figure}[ht!]
 \includegraphics[width=5.0in]{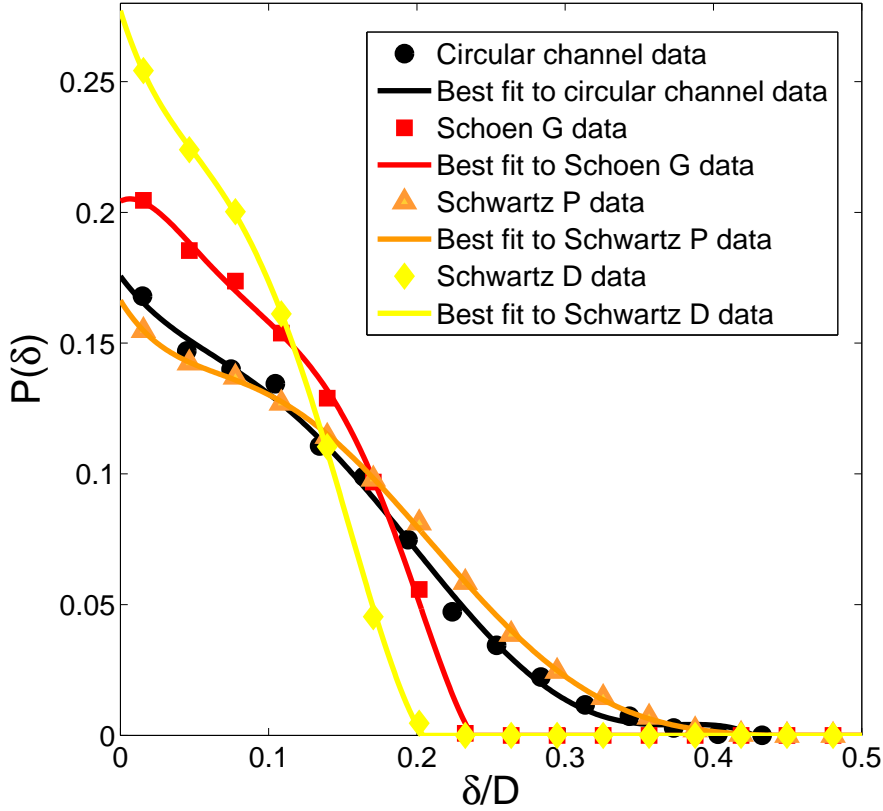}
 \caption{\label{fig:pore_size} (Color online) Pore size density 
 function data generated from the presented algorithm for the three 
 minimal surfaces and the circular-channel model, along with the 
 best-fit fifth degree polynomial.}
\end{figure}

We employ the following algorithm to 
determine the pore-size function $P(\delta)$~\cite{coker95}:

\begin{enumerate}
 \item Choose a random location in the pore phase.
 \item Find the radius of the largest sphere centered at the above 
 point that just touches the two-phase interface.  
 \item Repeat steps 1-2 for many random locations and create a 
 list of radii.
 \item After sampling sufficiently, bin the sphere radii.  Divide 
 the number of radii in each bin by the total number of radii to 
 determine $P(\delta)$.  
\end{enumerate} 
The interpolated functional form of $P(\delta)$ is given in Eq. 
(\ref{SchwartzP}), (\ref{SchwartzD}), (\ref{SchoenG}) and 
(\ref{circ_channel}) for the Schwartz P, Schwartz D, Schoen G minimal 
surfaces, and the circular channel model (spherical pore circular-channel 
model with $b$ = 0), respectively [see Fig.~\ref{fig:pore_size}]:
\begin{eqnarray} 
  P(\delta) = \left\{ \begin{array}{cr}
  -149\delta^5 + 166\delta^4 - 61.2\delta^3 + 8.2\delta^2 - 
  0.696\delta + 0.164, & \textrm{if $\delta < 0.419$} \\
   0, & \textrm{otherwise} 
 \end{array} \right.
 \label{SchwartzP}
\end{eqnarray}
 
\begin{eqnarray} 
  P(\delta) = \left\{ \begin{array}{cr}
   2238\delta^5 - 523\delta^4 - 44.1\delta^3 + 12.8\delta^2 - 
   1.55\delta + 0.275, & \textrm{if $\delta < 0.202$} \\
   0,  & \textrm{otherwise} 
 \end{array} \right.
 \label{SchwartzD}
\end{eqnarray}
 
\begin{eqnarray} 
  P(\delta) = \left\{ \begin{array}{cr}
   -1388\delta^5 + 897\delta^4 - 225\delta^3 + 23.4\delta^2 - 
   1.47\delta + 0.223, & \textrm{if $\delta < 0.234$} \\
   0, & \textrm{otherwise} 
 \end{array} \right.
 \label{SchoenG}
\end{eqnarray}
  
\begin{eqnarray} 
  P(\delta) = \left\{ \begin{array}{cr}
   -162\delta^5 + 173\delta^4 - 59.9\delta^3 + 7.57\delta^2 - 
   0.764\delta + 0.176, & \textrm{if $\delta < 0.234$} \\
   0,  & \textrm{otherwise}. 
 \end{array} \right.
 \label{circ_channel}
\end{eqnarray}

\noindent For the spherical pore (spherical pore circular-channel 
model with $a$ = 0), the pore size density function is known exactly:

\begin{eqnarray} 
  P(\delta) = \left\{ \begin{array}{cr}
   \frac{\a 3(a-\delta)^2}{\a a^3}, & \textrm{if $\delta < a$} \\
   0,  & \textrm{otherwise.} 
 \end{array} \right.
\end{eqnarray}

\subsection{Universal Curve}

Based on rigorous lower bounds on the survival time \cite{torquato02}, the following ``universal" scaling relation 
for $\tau$ has been found to apply to a wide class of microstructures 
and range of porosities~\cite{torquato97}:
\begin{equation}
 \frac{\a \tau}{\tau_0} = \frac{\a 8}{\a 5}x + \frac{\a 8}{\a 7}x^2,
 \label{universal}
\end{equation}
where $\tau_0 = 3\phi_2/(D\phi s^2)$,  
$x = \langle \delta \rangle ^2/(\tau_0 D)$, and 
$\langle \delta \rangle$ is the mean pore size, defined by the
first moment of the pore-size density function 
$P(\delta)$~\cite{torquato02}:
\begin{equation}
 \langle \delta \rangle = \int_0^\infty \delta P(\delta)\;d\delta.
\label{pore}
\end{equation}

\begin{figure}[ht!]
 \includegraphics[width=4.5in]{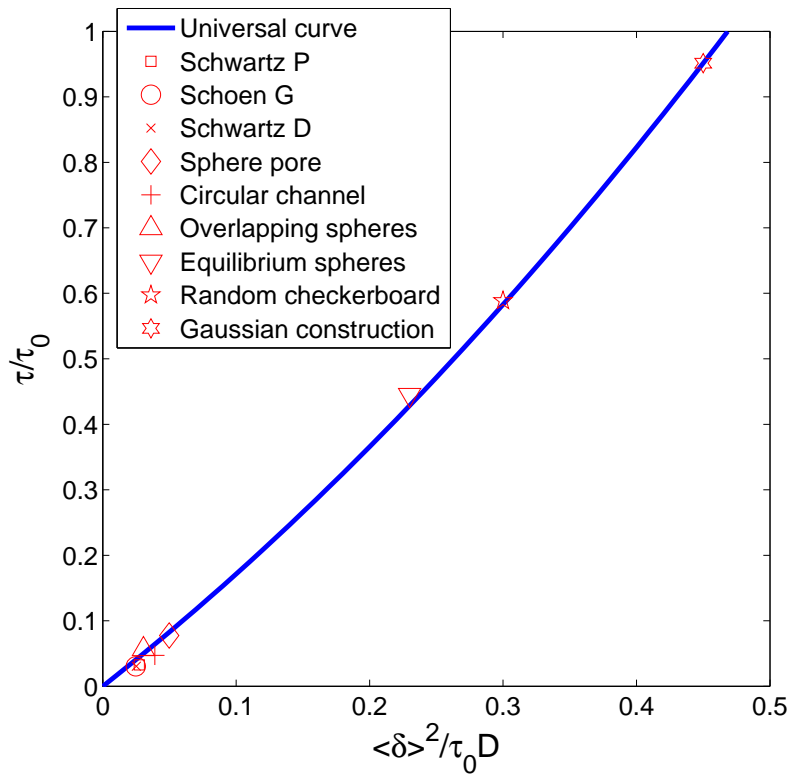}
 \caption{\label{fig:universal} (Color online) The dimensionless mean 
 survival time  $\tau/\tau_0$ versus  the scaled mean pore size squared,
$\langle \delta \rangle^2/(\tau_0 D)$,  for various two-phase media.  
 The solid curve is the universal scaling relation (\ref{universal}) \cite{torquato97}.  }
\end{figure}

To explore the robustness of the universal curve, we check if the 
reported mean survival time data for a subset of the triply periodic 
surfaces falls on the curve. In order to compare the reported $\tau$ 
results to Eq. (\ref{universal}), we determine 
$\langle \delta \rangle$ from the pore-size density function
$P(\delta)$ results given above.
Figure~\ref{fig:universal} reveals that the scaled survival times of the 
minimal surfaces fall relatively close to the universal curve. The 
corresponding values for the other bicontinuous structures considered 
here are close to the minimal-surface values and hence are not shown 
in the figure.

\section{CONCLUSIONS and DISCUSSION}

To conclude, we have shown that a simply connected pore phase is 
a crucial topological feature required 
to achieve large mean survival times at a porosity $\phi=1/2$. However, 
a simply connected pore phase is not a sufficient condition feature; one 
must also have interfaces with small specific surface $s$. Indeed, we 
found that the survival times are inversely proportional to the 
corresponding specific surfaces for all of the triply periodic 
structures considered here.  Nonetheless, the pore spaces of structures 
with large survival times are expected to be simply connected and 
therefore it would not be unreasonable for $\tau$ to be inversely
proportional to the specific surface in these instances.  We have 
demonstrated,  as hypothesized, that the Schwartz P porous medium 
maximizes the mean survival time among the set of twelve triply-periodic 
structures considered here.  This lends further evidence to the 
multifunctional optimality of the Schwartz P minimal surface, making 
this structure of great practical value to guide the design of new 
materials with a host of desirable bulk properties. Moreover,
scaled survival times of the minimal surfaces fall relatively close to the 
universal curve (\ref{universal}) found in Ref. \cite{torquato97}.

Based on our findings, we conjecture that the mean survival time (like 
the fluid permeability) is maximized for triply-periodic porous media with 
a simply connected pore space at porosity $\phi=1/2$ by the structure the 
globally optimizes the specific surface. The verification of this 
conjecture remains an outstanding open question. This extremal problem 
falls in the general class of {\it isoperimetric problems}, which are 
notoriously difficult to solve. A prototypical isoperimetric example is 
Kelvin's problem: the determination of the space-filling arrangement of 
closed cells of equal volume that minimizes the surface area. Although it 
is believed that the Weaire-Phelan structure \cite{We94} is an excellent 
solution to Kelvin's problem, there is no proof that it is a globally 
optimal one. Our conjecture is also likely a  difficult one to prove. 

In this regard, it is noteworthy that an original goal of Ref. 
\cite{jung07} was to show that the triply-periodic surface with minimal 
specific surface $s$ at porosity $\phi= 1/2$ is the Schwartz P surface. 
While numerical simulations provided empirical evidence supporting this 
proposition, the authors of Ref. \cite{jung07} could not prove it 
rigorously. However, they were able to show that the Schwartz P, Schwartz D, 
and Schoen G minimal surfaces are local minima of the specific surface area 
$s$ at fixed volume fraction $\phi=1/2$. Thus, the question of the global 
optimality of the Schwartz P surface (i.e., minimal total interface surface 
area or specific surface $s$), is an open question for future investigation.

\section*{ACKNOWLEDGMENTS} 

S. T. thanks the Institute for Advanced Study for its hospitality during 
his stay there. This work was supported by the Office of Basic Energy 
Sciences, U.S. Department of Energy, under Grant No. DE-FG02-04-ER46108.


\begin{thebibliography}{11}
\bibitem{compet}
Specifically, these are
the optimal structures (at least to an excellent approximation \cite{Si07})
when a weighted sum of the effective thermal and electrical conductivities is maximized for the
case in which phase 1 is a good thermal conductor but a poor
electrical conductor and phase 2 is a poor thermal conductor
but a good electrical conductor.
The demand that this sum is maximized sets up a competition between the two effective
transport properties, and this demand is met by the Schwartz
P and D structures.
We note that for reasons of mathematical
analogy, the optimality of these bicontinuous composites applies
to any of the pair of the following scalar effective properties:
electrical conductivity, thermal conductivity, dielectric
constant, magnetic permeability, and diffusion coefficient.
\bibitem{Si07}
L. Silvestre, Proc. R. Soc. A {\bf 463}, 2543 (2007).
This paper suggests that the minimal surfaces
are strictly not optimal, even if they are to an excellent
approximation. However, as pointed out by this author,
this distinction (even if true) is not relevant for practical considerations.
Thus, we call them extremal, since they are for all practical purposes.
\bibitem{simply}
A region of space is said to be simply connected if any simple closed surface 
can be shrunk to a point continuously in the region
(i.e., a surface that has the same topology as a sphere).
Therefore, a simply connected pore region cannot have
any ``holes" or ``dead ends."
\end{thebibliography}
\end{document}